\documentstyle[prb,aps,epsf]{revtex}
\begin{document}

\twocolumn[\hsize\textwidth\columnwidth\hsize\csname@twocolumnfalse%
\endcsname

\title{Propagation failure of traveling waves
in a discrete bistable medium}

\author{G\'abor F\'ath\cite{byline1}}
\address{Institute of Theoretical Physics, University of Lausanne,\\
         CH-1015 Lausanne, Switzerland}

\date{\today}

\maketitle

\begin{abstract}
Propagation failure (pinning) of traveling waves is
studied in a discrete scalar reaction-diffusion equation with a
piecewise linear, bistable reaction function.
The critical points of the pinning transition, and the wavefront profile
at the onset of propagation are calculated exactly. The scaling of the
wave speed near the transition, and the leading corrections to the front
shape are also determined. We find that the speed vanishes logarithmically
close to the critical point, thus the model belongs to a different
universality class than the standard Nagumo model, defined with a smooth,
polynomial reaction function.
\end{abstract}

\vskip 0.3 truein
]

\section{Introduction}

Reaction-diffusion equations possessing
traveling wave solutions have wide applications in various fields,
from chemistry and biology to physics and
engineering.\cite{books} Even though the underlying
physicals systems are often composed of discrete elements, such
as biological cells, the continuum approximation,
described by partial differential equations, usually works well for 
realistic model parameters. However, there are important exceptions
when the intrinsic discrete character of the medium manifests itself,
and the resulting phenomena cannot be adequately described without
incorporating the discreteness into the mathematical model.

A typical example, when spatial discreteness plays a crucial role,
is the phenomenon of {\em propagation failure} or {\em pinning} of
traveling waves.
Consider the scalar reaction-diffusion equation (Nagumo equation)
\begin{equation}
\frac{\partial u}{\partial t}  =
D\frac{\partial^2 u}{\partial x^2} + f(u),
\label{rd_cont}
\end{equation}
where $u=u(x,t)$ is the chemical concentration function, and $D$ is the
diffusion constant. The reaction function $f(u)$ is assumed
to be bistable (Nagumo-type), and depends on a parameter $a$, $0<a<1$,
such that
\[
f(0)=f(a)=f(1)=0,
\]
and
\[ 
\left. \begin{array}{ll}
              f(u)<0\quad  & \mbox{if $\;0<u<a$,}\\
              f(u)>0\quad  & \mbox{if $\;a<u<1$}.
              \end{array}
       \right.
\] 
It is known\cite{Aron} that whenever $\int_0^1 f(u)du\ne 0$,
Eq.\ (\ref{rd_cont}) possesses a unique, stable traveling wave
solution of the form
\begin{equation}
u(x,t)=U(z),\qquad z=x-ct,
\label{TW_form}
\end{equation}
for any value of $D$. The diffusion constant, which can be
scaled to unity by rescaling $x$, only influences the wave
speed $c$ in a trivial manner 
\begin{equation}
c=c_0 D^{1/2},
\label{c_cont}
\end{equation}
where $c_0$ is independent of $D$.

The {\em discrete} analog of Eq.\ (\ref{rd_cont}) is defined as
\begin{equation}
\frac{\partial u_n}{\partial t}  =
D (u_{n-1}+u_{n+1}-2u_n)+ f(u_n).
\label{rd_disc}
\end{equation}
Now the concentration function $u_n$ takes values only on a
one-dimensional lattice $n\in Z$. In contrast to the
continuum case, traveling wave solutions only
exist if the coupling between the lattice sites is strong enough. For
weak diffusion, propagation is impeded and, instead, a series
of stable steady state solutions emerge. The critical value $D_c$
where the pinning transition occurs depends on the actual reaction
function $f(u)$ and is not universal. However, it is believed that the
type of the transition, i.e., the way $c$ tends to zero as $D\to D_c^+$
(from above) is universal, and only depends on some global properties
of $f(u)$.

The phenomenon of propagation failure in bistable systems have been
intensively studied in the last decade by several authors:
Keener\cite{Keen} studied propagation failure in the discrete
Nagumo equation and in some of its generalizations, like the
FitzHugh-Nagumo or the Beeler-Reuter 
models, in the context of stimulus conduction in the miocardial
tissue. He demonstrated that for weak coupling no propagation is
possible, and found lower and upper bounds for the critical coupling
$D_c=D_c(a)$ as a function of the reaction function parameter $a$.
He also showed using perturbation theory that in the large $D$
limit the leading correction to the
wave speed is an effective shift in $D$ so that  
\begin{equation}
c=c_0 (D-K)^{1/2},
\label{c_disc}
\end{equation}
where the constant $K$ can be obtained by quadrature from the (known)
solution of the associated continuum problem in Eq.\ (\ref{rd_cont}).

A rigorous proof that for strong enough coupling $D$ Eq.\
(\ref{rd_disc}) indeed has stable traveling wave solutions
in the form of Eq.\ (\ref{TW_form}), with $x$
replaced by $n$, was later given by Zinner.\cite{Zinn}

More recently, Erneux and Nicolis\cite{ErNi} suggested a general theory
for the critical behavior near the pinning transition. Using
series expansion techniques and bifurcation theory arguments
they concluded that as the transition
point is approached $c$ must vanish according to a power law
\begin{equation}
c\sim (D-D_c)^{1/2},\qquad D\to D_c^+
\label{c_crit}
\end{equation}
for $f(u)$ fixed. Of course, despite the formal similarity of Eqs.\
(\ref{c_disc}) and (\ref{c_crit}), the transition point $D_c$
cannot be identified with the constant $K$ of Eq.\ (\ref{c_disc}),
even though their numerical values may be close.\cite{PePeCh92}

The above theory of wave propagation and its failure
in discrete media has also been fruitfully applied to understand
experimental observations in linear arrays of continuously stirred tank
reactors,\cite{Lapl} coupled Chua circuits,\cite{PePeCh92,Chua} and in
the theory of excitable cables.\cite{Rinz} Applications to
morphogenetic pattern formation and cellular differentiation are
also under investigation.\cite{Fath}

In this paper we analyze the phenomenon of propagation failure
for the piecewise linear reaction function
\begin{equation}
f(u)=\left\{ \begin{array}{ll}
              -u      & \mbox{if $u\le a$}\\
              1-u  & \mbox{if $u> a$}.
              \end{array}
       \right.
\label{Flin}
\end{equation}
This is a case when the $a$-dependence of the critical coupling $D_c$
can be analytically found and the pinning transition can be directly
studied by relatively simple mathematical tools. Our principal aim is
to investigate the critical behavior of the system around $D_c$,
and derive the scaling law of the wave speed. At the same time,
we obtain the leading corrections to the wavefront profile $U(z)$. 

The organization of the paper is as follows: Section II gives an
overview of the traveling wave solutions in the continuum case. Section
III treats the discrete model, and consists of five subsections discussing
(A) the steady state solutions, (B) their stability, (C) the front shape
at the transition point, (D) the leading corrections to the front
shape for slow waves, and (E) the scaling law of the speed
close to the transition point. Finally, Sec.\ IV summarizes and discusses
the main results.

\section{Continuous model}

In this section we look for traveling wave
solutions of Eq.\ (\ref{rd_cont}) defined with the reaction function 
Eq.\ (\ref{Flin}). 
Substituting the form of Eq.\ (\ref{TW_form}) into
Eq.\ (\ref{rd_cont}), we obtain the ordinary differential equation
\begin{equation}
DU''+cU'+f(U)=0.
\label{eq2}
\end{equation}
We impose the boundary conditions
\begin{equation}
U(-\infty)=1\quad U(\infty)=0\quad U(0)=a,
\label{bc}
\end{equation}
and suppose that the wavefront $U(z)$ is monotonically decreasing. Thus
Eq.\ (\ref{eq2}) can be written as
\begin{equation}
DU''+cU'-U+\Theta(-z)=0,
\label{Uthetaz}
\end{equation}
where $\Theta(z)$ is the conventional step function. To solve this
equation we use the Fourier transform technique, which allows a direct
generalization to the discrete case later in Sec.\ III C. 

Introducing the Fourier transform
\begin{equation}
V(k)={\cal F}[U(z)]=\int_{-\infty}^\infty U(z)e^{ikz}dz
\label{Fouriertr}
\end{equation}
Eq.\ (\ref{Uthetaz}) transforms into 
\[ 
-k^2 D V(k) -ick V(k) -V(k) -\frac{i}{k} +\pi\delta(k) =0.
\] 
This can be solved for $V(k)$ yielding
\[ 
V(k) = V_1(k)+V_2(k),
\] 
with
\begin{eqnarray}
V_1(k) &=& \frac{-i}{k(ick+Dk^2+1)},\nonumber\\
V_2(k) &=& \frac{\pi\delta(k)}{ick+Dk^2+1}.
\label{V_cont}
\end{eqnarray}
Note that $V_1(k)$ in Eq.\ (\ref{V_cont}) has three simple poles at
$k=0,i\kappa_+,i\kappa_-$ with
\begin{equation}
\kappa_\pm = \frac{1}{2D}\left(-c\pm \sqrt{c^2+4D} \right),
\label{kappapm}
\end{equation}
thus the inverse transformation is straightforward. We obtain
\begin{eqnarray}
{\cal F}^{-1}[V_1(k)]
  &=& \frac{1}{2}-\Theta(z)
-\frac{e^{\kappa_+ z}\Theta(-z)}{D\kappa_+(\kappa_+-\kappa_-)}-\nonumber\\
  && \frac{e^{\kappa_- z}\Theta(z)}{D\kappa_-(\kappa_+-\kappa_-)},\\
{\cal F}^{-1}[V_2(k)]
  &=& \frac{1}{2},
\end{eqnarray}
which leads to 
\begin{eqnarray}
U(z) 
   &=& -\frac{e^{\kappa_+ z}\Theta(-z)}{D\kappa_+(\kappa_+-\kappa_-)}
       -\frac{e^{\kappa_- z}\Theta(z)}{D\kappa_-(\kappa_+-\kappa_-)}+
        \Theta(-z).
\label{UFour} \nonumber\\       
\end{eqnarray}
The continuity of $U(z)$ and its derivative at
the origin is automatically satisfied by this expression, irrespective of
the value of $c$. The second derivative, however, has a jump
discontinuity at $z=0$.

\begin{figure}[hbt]
\epsfxsize=\columnwidth\epsfbox{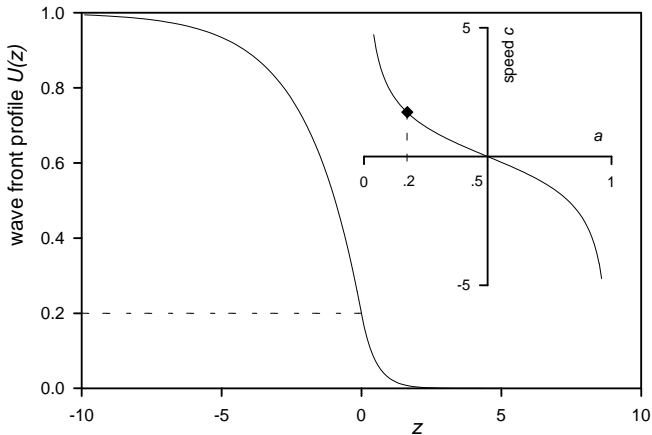}
\caption{Wave front profile $U(z)$ in the continuum case for the
diffusion constant $D=1$ and the reaction function parameter
$a=0.2$. The associated wave speed is $c=1.5$. The inset shows the wave
speed for general $a$.} 
\label{fig:cont} 
\end{figure}

The wave speed can be obtained from the
requirement $U(0)=a$ in Eq.\ (\ref{bc}). This gives the constraint
\[
\frac{-1}{D\kappa_-(\kappa_+-\kappa_-)}=a,
\]
from which $c$ turns out to be 
\begin{equation}
c=(1-2a)\sqrt{\frac{D}{a(1-a)}}.
\label{eq11}
\end{equation}
Note that this expression indeed satisfies the general form Eq.\
(\ref{c_cont}). We find that the speed changes sign at $a=1/2$,
and around this value it behaves linearly. Furthermore, $c$ diverges
whenever $a\to 0$ or $1$. Using Eq.\ (\ref{eq11}), the wave
front profile in Eq.\ (\ref{UFour}) reduces to
\begin{equation}
U(z)=\left\{ \begin{array}{ll}
              1-(1-a)e^{\kappa_+ z}  & \mbox{if $z\le 0$}\\
              ae^{\kappa_- z}  & \mbox{if $z\ge 0$},
              \end{array}
       \right.
\label{Uzfinal}
\end{equation}
where $\kappa_\pm$ is
\[
\kappa_+ = \sqrt{\frac{a}{(1-a)D}},\qquad
\kappa_- = -\sqrt{\frac{1-a}{aD}}.
\]
The wave front profile for $a=0.2$, and the $a$-dependence
of the speed for the continuum model is shown in Fig.\ \ref{fig:cont}.

\section{Discrete model}

We now turn to the analysis of the discrete problem defined by Eq.\
(\ref{rd_disc}) and (\ref{Flin}). 

\subsection{Steady states and virtual steady states}

The critical curves of the pinning transition on the $(D,a)$ parameter 
plane can be obtained by
analyzing the range of existence of monotonic steady state
solutions.\cite{Keen} We investigate under what conditions a monotonically
decreasing steady state
can exist with the boundary conditions $u_{-\infty}=1$,
$u_\infty=0$. Clearly, if such a solution exists, due to translational
invariance, an infinite series of solutions can be constructed. In the
following, any distribution $u_n^M$ of the concentration values
that satisfies the inequalities
\begin{equation}
\left. \begin{array}{ll}
              u^M_{n_{}} > a  & \mbox{if $n< M$}\\
              u^M_n \le   a  & \mbox{if $n\ge M$,}
              \end{array}
       \right. 
\label{consist}
\end{equation}
will be referred to as the {\em kink}-$M$ distribution (solution).
We introduce the superscript $M$ to explicitly denote the
position of the kink, lying between sites $M-1$ and $M$.
Such kink distributions contain two different
domains. The sites on the left (right) side of the kink are subject to
the high (low) concentration branch of the piecewise linear $f(u)$
function.

As is shown by Eq.\ (\ref{eq11}), the continuum problem
possesses steady state solutions (traveling wave solutions with $c=0$)
for a given $D$ only when $a=1/2$. In contrast, in the discrete
version, monotonic steady state kinks persist over a wide range of
values of $a$. The calculation of this range as a function of $D$
for the piecewise linear reaction function in Eq.\ (\ref{Flin}) is
straightforward: Providing that
Eq.\ (\ref{consist}) holds, a candidate kink-$M$
{\em steady state\/} solution $u_n(t)=s_n^M$ can be looked for in the form
\begin{equation}
s^M_n =  \left\{ \begin{array}{ll}
              1-A_+ e^{\kappa n} & \mbox{if $n< M$}\\
              A_- e^{-\kappa n}  & \mbox{if $n\ge M$.}
              \end{array}
       \right. 
\label{bulkkink}
\end{equation}
Substituting this expression into Eq.\ (\ref{rd_disc}), the inverse
diffusion length $\kappa$ reads
\begin{equation}
\kappa = \cosh^{-1}\left( \frac{1}{2D}+1 \right),
\label{kappa}
\end{equation}
while the two constants $A_+$ and $A_-$ turn out to be
\begin{equation}
A_\pm = a_\pm e^{\mp\kappa M},
\label{constA}
\end{equation}
with the $M$-independent part defined by
\begin{equation}
a_\pm = \frac{1}{2}\left(1\pm \frac{1}{\sqrt{1+4D}}\right).
\label{aunder}
\end{equation}

Note, however, that $s_n^M$ in Eq.\ (\ref{bulkkink}) is a valid
kink-$M$ steady state only if it satisfies the inequalities in
Eq.\ (\ref{consist}). Since $s_n^M$ is monotonically decreasing, the
conditions $s^M_{M-1}>a$ and $s^M_M\le a$ are necessary and sufficient.
From Eq.\ (\ref{bulkkink}) we find that
\[ 
s^M_{M-1}=a_+,\qquad s^M_M=a_-,
\] 
thus we conclude that monotonically decreasing steady states exist for
a given $D$ only if $a_-<a<a_+$, where $a_\pm=a_\pm(D)$ is expressed
by Eq.\ (\ref{aunder}). We point out that the shape of the steady state,
when it exists, do not depend on the actual parameter $a$, but only on
the diffusion constant $D$. 

\begin{figure}[hbt]
\epsfxsize=\columnwidth\epsfbox{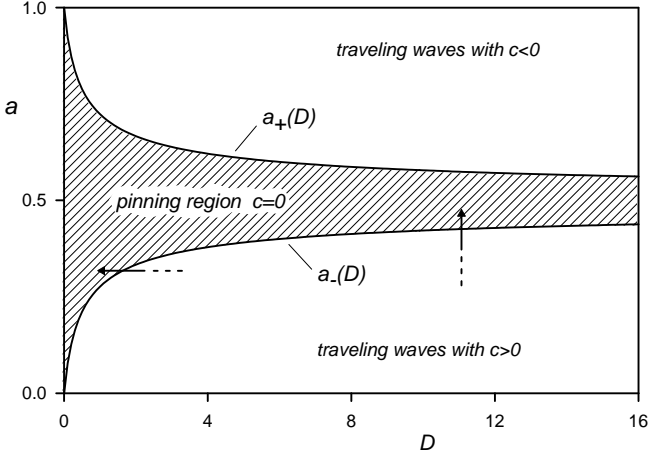}
\caption{Phase diagram on the $a$ vs $D$ plane. Pinned steady state
solutions exist in the shaded region, while traveling waves exist
in the unshaded ones. The pinning transition takes place along
the $a=a_+(D)$ and the $a=a_-(D)$ curves, and can be initiated either by
changing $D$ for $a$ fixed, or by varying $a$ with $D$ fixed (see
arrows).} 
\label{fig:pin_region} 
\end{figure}

The values of $a_-=a_-(D)$ and $a_+=a_+(D)$ are plotted in Fig.\
\ref{fig:pin_region} as a function of the diffusion constant $D$.
As was argued by Keener\cite{Keen}, the existence of steady state
solutions impede propagation, thus traveling wave solutions do not
exist in the shaded region (pinning region) of the figure. They do exist,
however, when the parameter pair $(D,a)$ lies outside the shaded regime,
i.e., if $a <a_-$ or $a >a_+$. In such cases, the concentration distribution
$s_n^M$ defined by Eqs.\ (\ref{bulkkink}--\ref{constA}) is not
a valid steady state. Thus, because of the role it plays in the
description of the dynamics of traveling waves, we will refer to it
in later sections as the {\em virtual} kink-$M$ steady state.

\subsection{Stability of the kink-M steady state}

To assess the stability of the kink-$M$ steady state, we introduce a
perturbation $\delta u_n(t)$ as
\begin{equation}
u_n(t)=s_n^M-\delta u_n(t).
\label{perturb}
\end{equation}
Substituting Eq.\ (\ref{perturb}) into Eq.\ (\ref{rd_disc}),
the dynamics of the perturbation is found to satisfy the equation
\begin{equation}
\frac{\partial \delta u_n}{\partial t}  =
D (\delta u_{n-1}+\delta u_{n+1}-2\delta u_n)-\delta u_n.
\label{delta_u}
\end{equation}
This is the discrete diffusion equation with an additional spatially
uniform decay term. Clearly, it only has decaying solutions, so
our kink-$M$ steady state is {\em stable}.

Nevertheless, for use in later sections, we investigate Eq.\
(\ref{delta_u}) further. Introducing the Fourier transform and the
Laplace transform of $\delta u_n(t)$ as
\begin{eqnarray*}
\delta u(k,t) &=& 
   \sum_{n=-\infty}^\infty \delta u_n(t)\, e^{ikn}, \\
\delta u_n(s) &=& 
   \int_0^\infty dt\, \delta u_n(t)\, e^{-st}, 
\end{eqnarray*}
respectively, we find
\[ 
\delta u(k,s) = \frac{\widetilde{\delta u}(k,0)}
                                {s+2D(1-\cos k)+1},
\] 
where $\widetilde{\delta u}(k,0)$ is the Fourier transform of the
initial perturbation at time $t=0$. Inverting the
transformations we obtain
\[ 
\delta u_n(t)=\int_{-\pi}^{\pi}\frac{dk}{2\pi}
            \widetilde{\delta u}(k,0)
            e^{-[1+2D(1-\cos k)]t}
            e^{-ikn}.
\] 
The fundamental solution (Green's function) of Eq.\ (\ref{delta_u}),
i.e., the dynamics for the initial condition 
$\delta u_n(0)=\delta_{n,m}$ turns out to be
\[ 
\delta u_n(t)=e^{-(1+2D)t} \int_{-\pi}^{\pi}\frac{dk}{2\pi}
                        e^{2Dt\cos k}
                        \cos[k(n-m)].
\] 
Noting that the definition of the modified Bessel function of integer
order $I_n(z)$ is\cite{Abra}
\[ 
I_n(z)=\frac{1}{\pi} \int_{0}^{\pi}
                        e^{z \cos \theta}
                        \cos(n \theta) d\theta,
\] 
we can write the fundamental solution as
\begin{equation}
\delta u_n(t)=e^{-(1+2D)t} I_{n-m}(2Dt).
\label{funda}
\end{equation}

\subsection{Front shape at c = 0$^+$}

The Fourier transform method can be used to determine the shape of the
propagating front $U_0(z)$ in the limit $c\to 0^+$. Introducing $z=n-ct$,
we look for solutions in the form
\[ 
u_n(t)=U(z).
\] 
The wave front profile function $U(z)$ satisfies the functional
differential equation
\begin{eqnarray}
D\left[U(z-1)+U(z+1)-2U(z)\right] +&& \\
cU'(z)-U(z)+\Theta[U(z)-a] &&= 0,
\label{eq:sing}
\end{eqnarray}
containing ``advanced'' and ``delayed'' terms of the continuous variable
$z$. Supposing that $U(0)=a$ and $U(z)$ is monotonically decreasing,
the last term can be rewritten as
\[ 
\Theta[U(z)-a]=\Theta(-z).
\] 

Fourier transforming, using the definition in Eq.\ (\ref{Fouriertr}), yields
\begin{equation}
2D(\cos k -1) V(k) -ick V(k) -V(k) -\frac{i}{k} +\pi\delta(k) =0,
\label{eqforV}
\end{equation}
which can be solved for $V(k)$ as
\[ 
V(k) = V_1(k)+V_2(k),
\] 
with
\begin{eqnarray}
V_1(k) &=& \frac{-i}{k[ikc+2D(1-\cos k)+1]}, \nonumber\\
V_2(k) &=& \frac{\pi\delta(k)}{ikc+2D(1-\cos k)+1}.
\label{Vdisc}
\end{eqnarray}
These expressions only differ from those in Eq.\ (\ref{V_cont}) in
that they contain $2(1-\cos k)$ instead of $k^2$.

As before, $V_2(k)$ only contributes as a uniform shift of $1/2$ to
$U(z)$. The inverse Fourier transform of the
$V_1(k)$ term is more subtle, since now it has an infinite series
of poles due to the different branches of the $\cos k$ function.
Explicit form of $U(z)$ can only be obtained in the limit $c=0^+$.

\begin{figure}[hbt]
\epsfxsize=\columnwidth\epsfbox{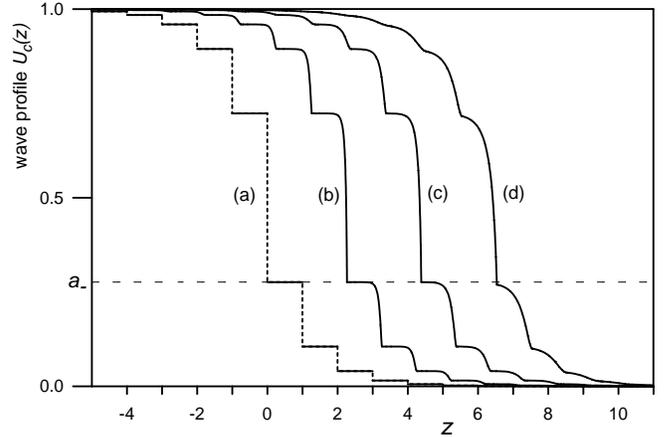}
\caption{Wave front profiles $U_c(z)$ near the pinning transition
for $D=1$. The corresponding exact critical point is
$a_-= 1/2-1/2 \protect\sqrt{5} =0.27639320\dots$.
The four curves belong to different parameters $a$ and
velocities $c$: (a) $a=a_-$, $c=0$, (b) $a=0.276393$,
$c=0.0780$, (c) $a=0.276$, $c=0.1782$, (d) $a=0.27$, $c=0.3246$.
Curves (b)--(d) are shifted for better visibility.}
\label{fig:plateaus} 
\end{figure}

When $c=0$, we use Eq.\ (\ref{kappa}) to write $V_1(k)$ as
\[ 
V_1(k) = \frac{i}{\sqrt{4D+1}}  \frac{1}{k}
           \left[\frac{1}{1-e^{ik+\kappa}}+
               \frac{1}{1-e^{ik-\kappa}}  \right].
\] 
After some trivial manipulations to assure convergence, the two terms
can be rewritten as infinite sums of geometrical series, leading to
\begin{eqnarray*}
V_1(k) &&= \frac{-i}{\sqrt{4D+1}} \frac{1}{k}
           \left[
             \sum_{n=1}^\infty e^{-ikn}e^{-\kappa n} +
             \sum_{n=0}^\infty e^{ikn}e^{-\kappa n}
           \right]\nonumber\\
&&       = \frac{-i}{\sqrt{4D+1}} \frac{1}{k}
           \left[
             1+
             2\sum_{n=1}^\infty \cos(k n)e^{-\kappa n}
           \right].
\end{eqnarray*}
According to mathematical handbooks,\cite{Erde}
\begin{eqnarray*}
&&{\cal F}^{-1}\left[\frac{-i}{k}\right] =
                  -\frac{1}{2}+\Theta(-z),\\
&&{\cal F}^{-1}\left[\frac{-2i}{k}\cos(k n)\right] =
                   \Theta(-n-z)-\Theta(z-n),
\end{eqnarray*}
thus, collecting all the terms, the final result for
$U_0(z)\equiv \lim_{c\to 0^+} U(z)$ is
\begin{eqnarray*}
U_0(z) =  a_- +&& \frac{1}{\sqrt{4D+1}}
   \left\{ \Theta(-z)+\frac{}{}  \right. 
   \nonumber\\
   && \sum_{n=1}^\infty
          \left[ \Theta(-n-z)-\Theta(z-n) \right]e^{-\kappa n}
   \left. \frac{}{} \right\}.
\end{eqnarray*}
This function is plotted in Fig.\ \ref{fig:plateaus} as curve (a).

As is seen, $U_0(z)$ contains a series of plateaus separated by jump
discontinuities. Consequently, the propagation of the front in the $c\to
0^+$ limit is characterized by an (infinitely) long {\em waiting}
period, when the concentrations hardly change, followed by
a rapid {\em jump} period, when the front steps ahead a lattice site.
We can calculate the values of $U_0(z)$ on the plateaus, i.e., in the
waiting period. We obtain that for any $z$ in the range $n<z<n+1$
\[ 
U_0(z) = \left\{ \begin{array}{ll}
              1-a_+ e^{\kappa n}  & \mbox{if $n< 0$}\\
              a_- e^{-\kappa n} & \mbox{if $n\ge 0$}.
              \end{array}
           \right. 
\] 
Comparing this formula to Eq.\ (\ref{bulkkink}), we see that the plateau
values coincide with the concentration values in the $a$-independent
steady state solution $s_n^0$ for the given value of $D$. Thus
close to the pinning transition, traveling
waves in their waiting period look almost identical to the associated
steady states belonging to the same $D$ but to a different $a$
parameter $a_-<a<a_+$.

\subsection{Leading correction to the front shape}

Our next aim is to calculate the leading correction to the shape of
the front when $0<c<<1$. When the front propagates with a nonzero speed,
the corresponding wavefront profile $U_c(z)$ changes drastically
in the narrow {\it transition layers} around the jump discontinuities of
$U_0(z)$, so that $U_c(z)$ becomes {\em continuous\/}. While the spacing
of these layers is unity (the lattice constant was set to unity),
their widths are proportional to $c$. In the limit $c\to 0^+$, all
the transition layers shrink to instantaneous jumps (see Fig.\
\ref{fig:plateaus}).

Note that the small extra term $cU^\prime(z)$ in Eq.\ (\ref{eq:sing})
is a {\it singular perturbation}. When it is absent, i.e., $c=0$,
the order of our functional differential equation Eq.\ (\ref{eq:sing}) is reduced.
The effect of this term on the
solution cannot be calculated by regular perturbation theory, i.e., series
expansion with respect to the small parameter $c$. There are regions of the
solution function, the transition layers, where $U_c(z)$ deviates in
an essentially nonperturbative way from the solution of the unperturbed
problem $U_0(z)$. The singular nature of the problem also manifests
itself in the Fourier transform $V_1(k)$ in Eq.\ (\ref{Vdisc}). Since
the poles are solutions of a transcendential equation, even the
slightest change in the term $ikc$ will substantially modify the pole
structure far from the origin. 

\begin{figure}[hbt]
\epsfxsize=\columnwidth\epsfbox{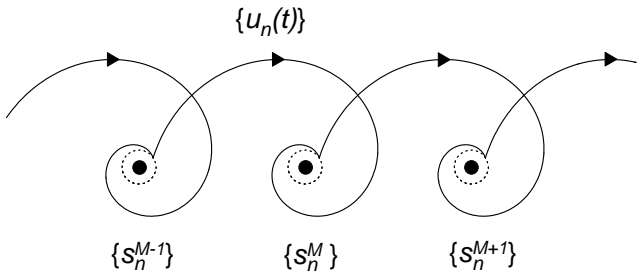}
\caption{Schematic illustration of the role of virtual steady states
$\{s_n^M\}$ in the dynamics. The continuous curve
represents the time evolution of the system in the infinite dimensional
vector space $\{u_n\}$. When a virtual attractor (black dot) is
approached as close as the dotted circle, the consistency condition
breaks down and a new stable-looking attractor emerges one lattice site
ahead.} 
\label{fig:attractors} 
\end{figure}

In order to obtain the functional form in the transition layers,
we return to Eq.\ (\ref{rd_disc}), and analyze its real time dynamics in a
small time period directly after the concentration at a
given site, say at site 0, has exceeded the value of $a$. At this
moment ($t=0$) the concentration values in the moving front are
distributed in a way that they satisfy the inequalities in
Eq.\ (\ref{consist}) with $M=1$, i.e., the kink is located between sites
0 and 1. We have seen in Sec.\ III A and B that
for such a situation a stable steady state solution exists,
namely $s_n^1$ defined by Eq.\ (\ref{bulkkink}). Therefore in these very
first moments the dynamics of 
$u_n(t)$ is identical to the gradual approach towards the stable-looking
distribution $s_n^1$.
However, since $(D,a)$ lies outside the pinning region, $s_n^1$ is only a
{\em virtual} steady state. The dynamical system will never attain
such a virtual attractor, because when the system approaches it
too closely, the value of $u_1(t)$ exceeds $a$ and the consistency
condition Eq.\ (\ref{consist}) breaks down. At that moment
($t=T$) a new
virtual steady state emerges one lattice site ahead, and the whole story
repeats itself; this time, from shifted initial conditions.
The role of the virtual attractors in the dynamics is schematically
illustrated in Fig.\ \ref{fig:attractors}.

To describe quantitatively the dynamics for $0<t<T$, we require the
initial distribution $u_n(0)$ at $t=0$. Supposing that the
front propagates very slowly, $c\equiv 1/T<<1$, we can use the
observation that the transition layers are very narrow. Consequently,
$u_n(0)$ is well approximated by $u_n(0)\approx s_n^0$. 
Defining the deviation function $\delta u_n(t)>0$ by
\[ 
u_n(t)=s_n^1-\delta u_n(t),
\] 
we conclude that in the time interval $0<t<T$ the dynamics of $\delta
u_n(t)$ is given exactly by Eq.\ (\ref{delta_u}) with the approximate
initial condition at $t=0$
\[ 
\delta u_n(0)=s_n^1-s_n^0=
    \frac{1}{\sqrt{1+4D}}\, e^{-\kappa |n|}.
\] 

\begin{figure}[hbt]
\epsfxsize=\columnwidth\epsfbox{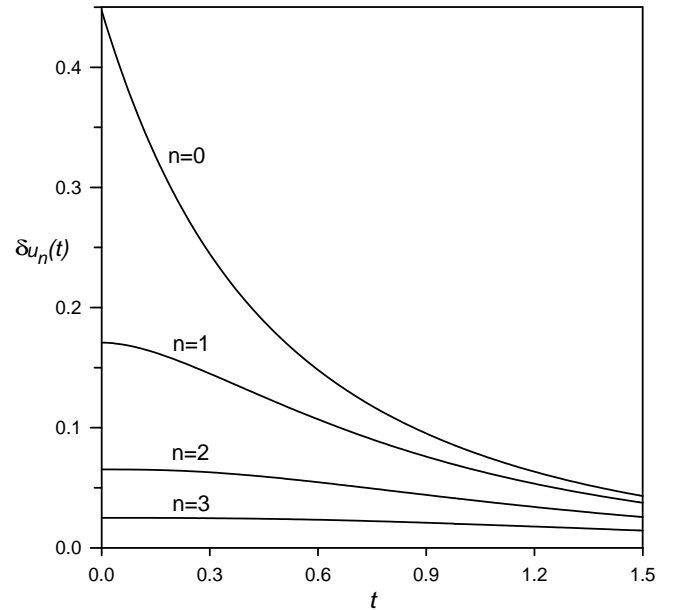}
\caption{The the deviation function $\delta u_n(t)$ is plotted vs
$t$ for $n=0,1,2$ and $3$. Except the $n=0$ curve, all the curves
start with zero slope at $t=0$.} 
\label{fig:deltau} 
\end{figure}

Using the fundamental solution in Eq.\ (\ref{funda}),
$\delta u_n(t)$ can be expressed as
\begin{equation}
\delta u_n(t)=
    \frac{e^{-(1+2D)t}}{\sqrt{1+4D}}\, \Sigma_n(2Dt),
\label{unSigma}    
\end{equation}
where $\Sigma_n(\tau)$ is defined as an infinite sum involving Bessel
functions 
\[ 
    \Sigma_n(\tau) \equiv 
    \sum_{m=-\infty}^{\infty} e^{-\kappa |m|}\,  I_{n-m}(\tau).
\] 
Unfortunately, $\Sigma_n(\tau)$ cannot be expressed in a closed form with
known functions, but it can be reduced to quadrature: Introducing the
function
\[ 
F(c,\tau) \equiv \sum_{m=0}^\infty c^m I_m(\tau),
\] 
and its integral representation\cite{Hans}
\begin{eqnarray*}
F(c,\tau)&=&\exp\left(\left[c+\frac{1}{c}\right]\frac{\tau}{2}\right)-\\
&&\frac{1}{c\tau}\exp\left(\frac{\tau}{2c}\right)
\int_0^\tau d\xi\;\xi\, \exp\left(\frac{-\xi^2}{2\tau c}\right)I_0(\xi),
\end{eqnarray*}
we obtain
\begin{eqnarray}    
\Sigma_n(\tau) &=&
    2\cosh(\kappa n)\, F(e^{-\kappa},\tau)-e^{\kappa n}\,I_0(\tau)-\nonumber\\
    && 2\sum_{m=1}^{n-1} \sinh[\kappa (n-m)]\, I_m(\tau).
\label{sigmaF}    
\end{eqnarray}
This expression only contains a finite sum of $n-1$ terms involving
Bessel functions. For large values of $n$ it can also be expressed in an
even more compact form using another special integral, but we will not
need that form here. Figure \ref{fig:deltau} shows
the deviation function $\delta u_n(t)$,
calculated from Eqs.\ (\ref{unSigma}) and (\ref{sigmaF}), for some small
values of $n$.

Having solved the dynamics for $\delta u_n(t)$ in Eq.\ (\ref{unSigma}),
we are ready to express
the leading correction to the wave front profile in the slow
propagation limit. Using the formal definition $\delta u_n(t)=0$ for
$t<0$, we write the continuous function $U_c(z)$ in the compact form
\begin{equation}
U_c(z)=U_0(z)-\sum_{n=-\infty}^\infty
                   \delta u_n\left(\frac{n-z}{c}\right),
\label{leadcorr}
\end{equation}
recalling that for each $n$ the correction term $\delta u_n[(z-n)/c]$ is
negligibly small everywhere, except in the narrow layer
$n-{\cal O}(c)<z<n$.


There are two limits in which the functional form of $\delta u_n(t)$,
and hence that of $U_c(z)$ simplifies.
For small $\tau$, the $F(c,\tau)$ function can be expanded as
\[ 
F(c,\tau) \simeq 1+\frac{c}{2}\tau+\frac{c^2+2}{8}\tau^2+
                 {\cal O}(\tau^3).
\] 
Using this and the standard small argument expansions of the Bessel
functions, we obtain from Eq.\ (\ref{sigmaF})
\[ 
\Sigma_n(\tau) \simeq
\left\{ \begin{array}{ll}
   1+e^{-\kappa}\,\tau +{\cal O}(\tau^2)]  & \mbox{if $n=0$}\\
   e^{-n\kappa}\left[1+\cosh(\kappa)\, \tau+{\cal O}(\tau^2)\right] &
                                          \mbox{if $|n|\ge 1$.}
              \end{array}
       \right.
\] 
Hence, using Eq.\ (\ref{kappa}), we find that for $n=0$ 
\[ 
\delta u_0(t) \simeq  \frac{1}{\sqrt{1+4D}} -t+{\cal O}(t^2),
\] 
and for $|n|\ge 1$
\[ 
\delta u_n(t) \simeq \frac{e^{-\kappa n}}{\sqrt{1+4D}} +{\cal O}(t^2).
\] 
It is seen that
only $\delta u_0(t)$ starts with a finite slope at $t=0$, while
for all $|n|\ge 1$ the slope of $\delta u_n(t)$ is zero (see Fig.\
\ref{fig:deltau}). Consequently,
the jump discontinuities of $U_0(z)$ round off everywhere; $U_c(z)$
becomes continuous and differentiable for all $z$, except at the
origin, defined by $U_c(0)=a$, where the function retains a cusp
(see Fig.\ \ref{fig:plateaus}).

\subsection{Speed scaling near the transition point}

The other limit that we investigate is the large $t$ limit.
Using the large argument expansion of
the Bessel functions\cite{Abra}
\[ 
I_m(\tau)\simeq \frac{e^\tau}{\sqrt{2\pi \tau}}
      \left[1 +{\cal O}\left(\frac{m^2}{\tau}\right)\right],
\] 
we can directly evaluate $\Sigma_n(\tau)$ in leading order as
\[ 
\Sigma_n(\tau)
\simeq \frac{e^{\tau}}{\sqrt{2\pi \tau}}\sum_{m=-\infty}^{\infty} 
                                   e^{-\kappa |m|}
=\frac{e^{\tau}}{\sqrt{2\pi \tau}}\coth\frac{\kappa}{2}.
\] 
Substituting this formula into Eq.\ (\ref{unSigma}), we obtain
for large $t$
\begin{equation}
\delta u_n(t) \simeq \frac{e^{-t}}{\sqrt{4\pi Dt}}
+ {\cal O}\left( \frac{e^{-t}}{t^{3/2}} \right).
\label{large_t}
\end{equation}
This result can be used to derive a formula for the wave speed
close to the pinning transition: As was discussed above, the next step
of the wavefront occurs at time $T$, when the value of $u_1(T)$ reaches
(from below) the actual value of $a$. We introduce the small
variable $\delta \theta=\delta \theta(D,a)$ which measures the
difference from the critical point, defined by
\[ 
\delta \theta\equiv a_- -a=
\frac{1}{2}\left(1- \frac{1}{\sqrt{1+4D}}\right) -a.
\] 
Obviously
\[ 
\delta \theta \simeq\left\{ \begin{array}{ll}
 -\delta a              & \mbox{for $D$ fixed,}\\
 (1+4D)^{-3/2}\delta D  & \mbox{for $a$ fixed.}
              \end{array}
       \right. 
\] 
Using Eq.\ (\ref{large_t}), we find that in leading order the time $T$,
required for $\delta u_1$ to decrease to the value of $\delta \theta$,
satisfies the equation
\begin{equation}
\delta \theta=  \frac{e^{-T}}{\sqrt{4\pi D\, T}}.
\label{testable}
\end{equation}
Thus the wave speed $c\equiv 1/T$ is
\begin{equation}
c=\frac{-1}{\ln (\sqrt{4\pi D}\; \delta \theta)}+
{\cal O}\left(\frac{\ln(\ln 1/\delta\theta)}{\ln^21/\delta\theta}
                                                           \right),
\label{asympt}                                                           
\end{equation}
i.e., it behaves {\em logarithmically} off the critical point.

\begin{figure}[hbt]
\epsfxsize=\columnwidth\epsfbox{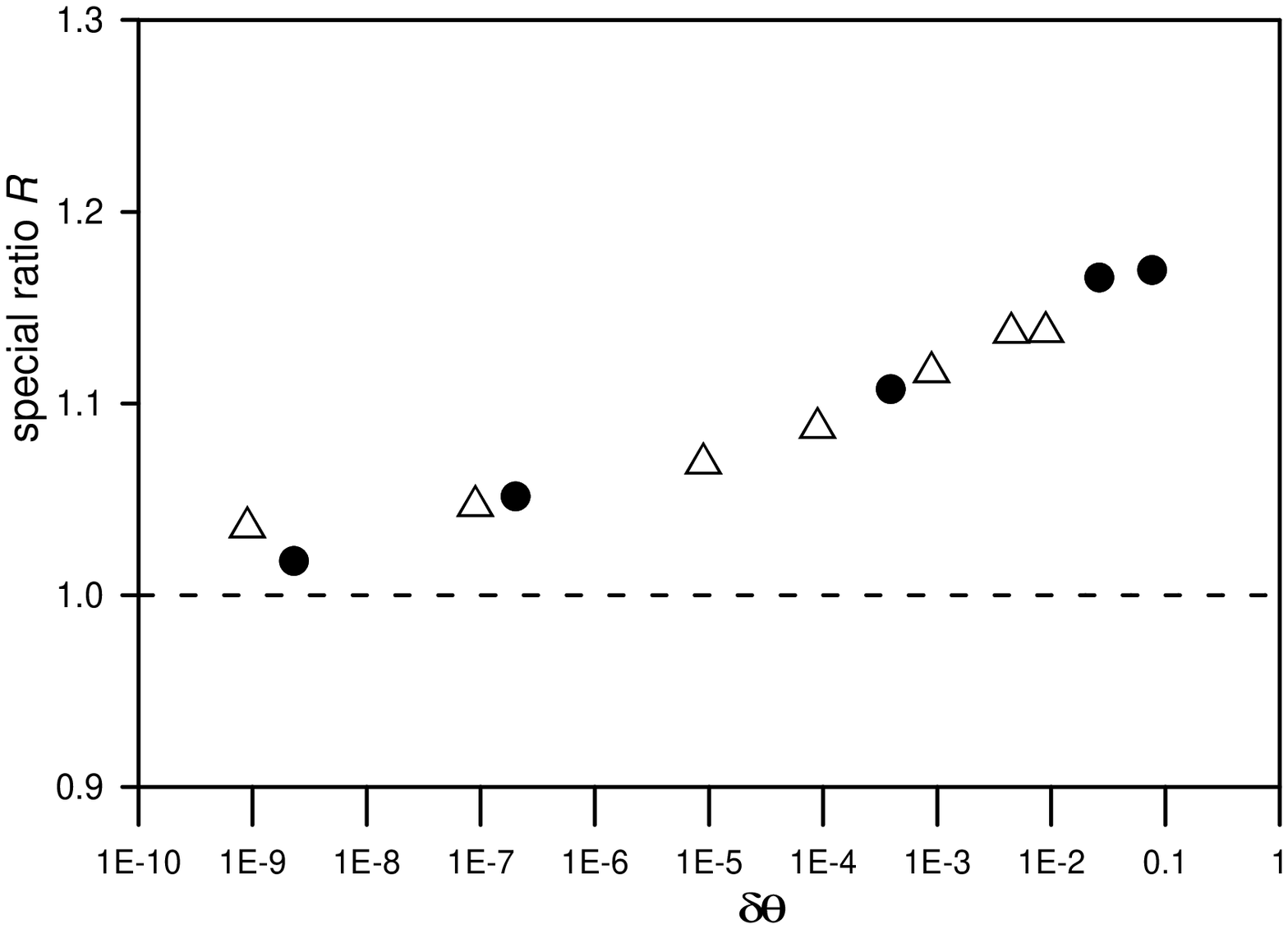}
\caption{The special ratio $R$, defined by Eq.\ (\protect\ref{R}),
is plotted vs the distance from the critical point $\delta\theta$
for the data obtained by numerical simulation. Circles refer to the
case when $D$ is varied, triangles to the one when $a$ is tuned around
the critical point $D=1$, $a=a_-(D)=1/2-1/2\protect\sqrt{5}$.
Theoretically $R$ approaches $1$ as $\delta\theta\to 0$.}
\label{fig:speed} 
\end{figure}

Even though the asymptotic form in Eq.\ (\ref{asympt}) is valid as
$\delta\theta\to 0$, we note that the log-log correction term is
significant for any numerically feasible values of $\delta\theta$. Thus
in order to test the scaling we use Eq.\ (\ref{testable}) instead, and
plot the numerically obtained values of
\begin{equation}
R=\frac{c^{1/2}e^{-1/c}}{\sqrt{4\pi D}\; \delta\theta}
\label{R}
\end{equation}
as a function of $\delta\theta$ in Fig.\ \ref{fig:speed}. As is seen,
the numerical values nicely approach the theoretical limit
$R=1$ as $\delta\theta\to 0$.

\section{Summary and discussion}

In summary, we have studied the phenomenon of propagation
failure in a particular discrete bistable reaction-diffusion model,
for which the reaction function $f(u)$ is composed of two stable
linear branches separated by a jump discontinuity at $u=a$.
This model is the simplest caricature of some widely used,
more realistic reaction functions such as the smooth polynomial
(Nagumo) function
\begin{equation}
f(u)=u(u-a)(1-u).
\label{polyf}
\end{equation}
So far, it was generally believed that the qualitative properties
of these models are identical.

In this paper we concentrated on the behavior of the piecewise linear
model near the pinning transition. The transition lines were
found by identifying the boundaries of the parameter domain
(pinning region) where steady state solutions exist.
By generalizing the concept of steady states to include virtual steady
states, we
showed that outside the pinning region the front propagates in a two
stage manner: an exponentially rapid ``jump-ahead'' period is followed
by a long ``waiting'' period when the concentrations hardly change. At
each moment
the system converges towards one of the stable-looking virtual
steady states, which is, however, never reached completely. We derived
analytic expressions for the wave front profile $U(z)$ in the slow
propagation limit, and argued that it becomes continuously differentiable
everywhere except at the origin, defined by $U(0)=a$, where the
function retains a cusp singularity.

We also calculated, in leading order, the wave speed as a function of the
distance from the critical point. Surprisingly, we found that
$c\to 0$ {\em logarithmically} as the transition point is approached.
This finding is distinct from previous results,\cite{ErNi}
obtained from the analysis of the reaction function in Eq.\
(\ref{polyf}), in which the speed was shown to scale as a
{\em power law} with an exponent $1/2$ in the pinning transition.

It is worth investigating whether the
different critical behavior found in our case is only due to the
discontinuity of our $f(u)$ function. To see this we analyzed
numerically the transition for the case when $f(u)$ 
is constructed from three linear pieces so that it remains continuous
everywhere, but possesses two nonanalytic break points where its
derivative jumps, 
\begin{equation}
f(u)=\left\{ \begin{array}{lll}
              -u \,  & \mbox{if $\, u\le a/2$}\\
              u-a\,  & \mbox{if $\, a/2\le u\le (a+1)/2$}\\
              1-u\,  & \mbox{if $\, u\ge (a+1)/2$}.
              \end{array}
       \right.
\label{3region}
\end{equation}
Our results (not shown) indicate that the
scaling of the speed near the transition point remains logarithmic
in this case, too.

It is clear that the analytic properties of the reaction functions
may play an important role in the critical behavior near the pinning
transition. It is less evident which singularities, in general,
are irrelevant and do not change universality, and which are
essential enough to modify the critical behavior,
as we found for the ones in Eqs.\ (\ref{Flin}) and (\ref{3region}).
The detailed analysis of these questions is left to future
investigation.


\acknowledgements
The author gratefully acknowledges valuable discussions with Z.\ Doma\'nski,
P.\ Erd\H os and S.\ B.\ Haley.
This research was supported by the Swiss National Science Foundation 
Grant No.\ 20-37642.93.

\end{document}